\documentclass{article}
\PassOptionsToPackage{table}{xcolor}
\PassOptionsToPackage{numbers}{natbib}
\usepackage{iclr2027_conference,times}

\usepackage{amsmath,amsfonts,bm}

\def\eqref#1{equation~\ref{#1}}

\def\1{\bm{1}}

\DeclareMathAlphabet{\mathsfit}{\encodingdefault}{\sfdefault}{m}{sl}
\SetMathAlphabet{\mathsfit}{bold}{\encodingdefault}{\sfdefault}{bx}{n}

\usepackage{amsmath,amssymb}
\usepackage{booktabs}
\usepackage{graphicx}
\usepackage{xcolor}
\usepackage{array}
\usepackage{tabularx}
\usepackage{placeins}
\usepackage{float}
\usepackage{enumitem}
\usepackage{microtype}
\usepackage{hyperref}
\usepackage{url}
\setcitestyle{authoryear,round}
\definecolor{ink}{HTML}{25313B}
\definecolor{muted}{HTML}{5F6B75}
\definecolor{accent}{HTML}{EAF3F0}
\definecolor{accentblue}{HTML}{EEF4F8}
\definecolor{rulegray}{HTML}{D7DEE3}
\definecolor{proposed}{HTML}{EAF4F7}
\definecolor{tablereference}{HTML}{DFF0F4}
\definecolor{tablediagnostic}{HTML}{F8E7C1}

\hypersetup{
  pdftitle={VoxReason: Auditing Source-Grounded Speech Plans Before Synthesis},
  pdfauthor={Mengzhe Geng},
  colorlinks=true,
  linkcolor=ink,
  citecolor=ink,
  urlcolor=ink
}

\setlist{nosep,leftmargin=*}

\newcolumntype{Y}{>{\raggedright\arraybackslash}X}
\newcommand{\repo}{\url{https://github.com/MENGZHEGENG/voxreason}}

\title{VoxReason: Auditing Source-Grounded\\ Speech Plans Before Synthesis}
\author{\textbf{Mengzhe Geng}\\
National Research Council Canada\\
\texttt{Mengzhe.Geng@nrc-cnrc.gc.ca}}
\date{}

\iclrfinalcopy

\begin{document}
\maketitle
\pagestyle{fancy}
\fancyhf{}
\fancyfoot[C]{\small\thepage}
\renewcommand{\headrulewidth}{0pt}
\renewcommand{\footrulewidth}{0pt}
\thispagestyle{fancy}

\begin{abstract}
Plan accuracy alone cannot show whether a speech-delivery decision follows its source: a fixed prior may match the original label yet fail to respond appropriately when a cue changes. VoxReason provides a 100-case verifier benchmark that holds each utterance fixed, edits one designated source-label cue, and scores cited evidence, eight plan fields, and the permitted response. On a source-key-disjoint test of 24 cases, a source-emotion prior reaches plan-slot accuracy $0.958\uparrow$, but none of the 24 edited neutral targets appears in its training labels; its required-change accuracy is $0.000\uparrow$. This diagnoses the support boundary of this prior, not its performance on supported edits. In a complementary 32-case emotion-disjoint test, the prior has seen all edited neutral targets but neither original test emotion; its plan-slot accuracy is $0.219\uparrow$ and required-change accuracy is $1.000\uparrow$. The partitions reuse and overlap the same 100 cases, so these deterministic diagnostics are not independent cohorts or learned-planner results. The benchmark evaluates derived labels and structured plans, not audio input, generated speech, or listener judgments.
\end{abstract}

\section{Introduction}

Speech generation systems are moving from fixed style controls toward context-aware decisions about delivery. A planner may need to infer whether a line should sound calm or urgent, how much pitch movement is appropriate, and which part of the context licenses that choice. Recent speech and audio language models connect language models to speech representations and generation tools \cite{speechgpt,audiogpt}, while newer work studies reasoning traces and context-aware text-to-speech \cite{thinksound,cottts}. These systems make the planning stage increasingly important: errors in the plan can be hidden by a fluent waveform, and a fluent waveform does not establish that the underlying decision was supported by the input.

The evaluation gap is specific. Existing audio benchmarks often ask for an answer about an audio clip or evaluate the quality of a generated waveform. Broad reasoning suites such as the benchmark in \cite{mmar} measure multi-step audio understanding across many domains. Codec and synthesis work likewise focuses on semantic preservation or output quality \cite{codec}. These settings are valuable, but they do not isolate whether a planner used the correct source evidence when the utterance is held constant and one contextual cue changes. Without that control, a system can exploit a stable text, speaker, or emotion prior and still appear context-aware.

VoxReason addresses this gap with an automatic pre-synthesis measurement protocol. It measures source attribution and plan consistency without listener ratings; it does not assess speech quality or listener preference. Each case contains a fixed target utterance, contextual metadata, source-label cues, an expected plan, and a counterfactual edit. The verifier checks cited cue identifiers, plan slots, citation coverage, and whether the edit changes only linked fields. Because the utterance remains fixed, the protocol isolates the structured planning decision from waveform quality. The repository is available at \repo\footnote{The repository contains derived labels, prompts, plans, and verification code. It does not redistribute the original RAVDESS audio files.}.

This paper makes three contributions. First, it defines a controlled task for source-grounded speech planning in which the utterance is fixed and one designated source cue changes. Second, it releases a compact source-label suite and an executable verifier with explicit metrics. Third, it checks whether the released task construction behaves as specified. It also tests shortcut predictors against a deterministic gold-copy reference. The package measures planning decisions; it does not evaluate listener judgments, waveform synthesis, or broad audio understanding.

\section{Related Work}

\paragraph{Speech and audio language models.}
SpeechGPT demonstrated a language-model architecture with cross-modal speech interaction \cite{speechgpt}, and AudioGPT organized audio understanding and generation through foundation-model tools \cite{audiogpt}. These systems motivate evaluation of intermediate decisions as multimodal capabilities expand. VoxReason is complementary to them: it does not introduce a new audio language model, and its primary test does not require waveform input. Instead, it asks whether a planner can ground a delivery decision in a supplied source record and update the decision when the record changes.

\paragraph{Reasoning benchmarks and structured generation.}
The benchmark in \cite{mmar} evaluates deep reasoning over speech, audio, music, and mixed content with a broad set of questions and reasoning layers. UniCATS uses surrounding context for speech continuation and editing \cite{unicats}. ThinkSound uses structured reasoning to guide audio generation and editing \cite{thinksound}, and the ISCSLP 2026 CoT-TTS challenge explicitly separates context understanding, reasoning output, and speech generation \cite{cottts}. VoxReason focuses on a narrower failure mode that these settings do not by themselves identify: unsupported or uncited evidence in a structured speech plan. The benchmark is consequently easier to inspect and less expressive than a full audio benchmark, but its intervention is more targeted.

\paragraph{Source data and acoustic representations.}
The case records are derived from the emotion and intensity labels of the Ryerson Audio-Visual Database of Emotional Speech and Song (RAVDESS) \cite{ravdess}. RAVDESS provides a controlled source for emotional speech, while VoxReason stores only derived labels, text prompts, source cues, and expected plans. The repository also includes an acoustic consistency check on extracted features; it is not a perceptual result. Work on semantic audio codecs highlights the importance of preserving meaning through acoustic representations \cite{codec}. VoxReason evaluates a stage before a codec or synthesizer is involved.

\section{Task Formulation}

Let a case be
\begin{equation}
  x = (u, r, s, E, p^{\star}, \Delta),
\end{equation}
where $u$ is the target utterance, $r$ is a role profile, $s$ is a scene label, $E$ is a set of source-label evidence records, $p^{\star}$ is the expected delivery plan, and $\Delta$ is a one-cue counterfactual edit. A planner receives the case input and returns
\begin{equation}
  \hat{y} = (\hat{C}, \hat{p}, \hat{q}),
\end{equation}
where $\hat{C}$ is a set of cited evidence records, $\hat{p}$ is a structured plan, and $\hat{q}$ is an optional rationale. The rationale is not scored as free-form prose; the verifier scores the structured evidence and plan fields.

The plan contains seven scalar fields---emotion, intent, pitch, energy, rate, pause, and stance---plus a set-valued emphasis field. A designated cue is the single source cue named by a counterfactual edit as the basis for changing linked plan fields. The edit changes that cue, such as an emotion or intensity label, while keeping the target text and other case fields fixed. The expected plan delta specifies which fields may change. The verifier evaluates three constraints:

\begin{enumerate}
  \item every cited record should be present in the case and match a gold evidence record;
  \item the plan should agree with the expected values across the eight scored slots; and
  \item after the edit, only the linked plan fields should change.
\end{enumerate}

This formulation distinguishes source grounding from surface plausibility. A system that emits the same plan for every cue may achieve high agreement on a biased split, but it fails the counterfactual locality test when the designated cue changes.

\section{VoxReason Benchmark and Verifier}

\subsection{Released source-label suite}

The public suite contains 100 cases with 67 training cases, 17 development cases, and 16 test cases. It has two target utterances, one scene label, and two role labels. The source labels yield 15 unique emotion-intensity keys, and all 15 keys map deterministically to a single expected plan in the released construction. This exact mapping makes the released split vulnerable to a source-emotion shortcut, which we test explicitly. Table~\ref{tab:profile} summarizes the released scope.

\begin{table}[t]
\centering
\caption{Public source-label benchmark profile. Counts are taken from the released files and their deterministic construct-validity checks.}
\label{tab:profile}
\small
\begin{tabularx}{0.82\linewidth}{@{}Yr@{}}
\toprule
{\bfseries\boldmath Property} & {\bfseries\boldmath Value}\\
\midrule
Total cases & 100\\
Train / development / test & 67 / 17 / 16\\
Target utterances & 2\\
Scene labels & 1\\
Role labels & 2\\
Emotion-intensity keys & 15\\
Deterministic key mappings & 15 / 15\\
Valid gold plans under prompt taxonomy & 100 / 100\\
Public context-audio records & 0\\
Source actors & 2\\
\bottomrule
\end{tabularx}
\end{table}

The suite is derived from RAVDESS source labels \cite{ravdess}. The public repository does not package the source audio. Consequently, the main experiment is a source-label measurement and not an audio-input benchmark. An acoustic feature summary and a matched comparison between source labels and extracted features describe consistency in the derived records. Neither evaluates listener perception or generated-speech quality.

\subsection{Verification path}

\begin{figure}[t]
\centering
\includegraphics[width=\linewidth]{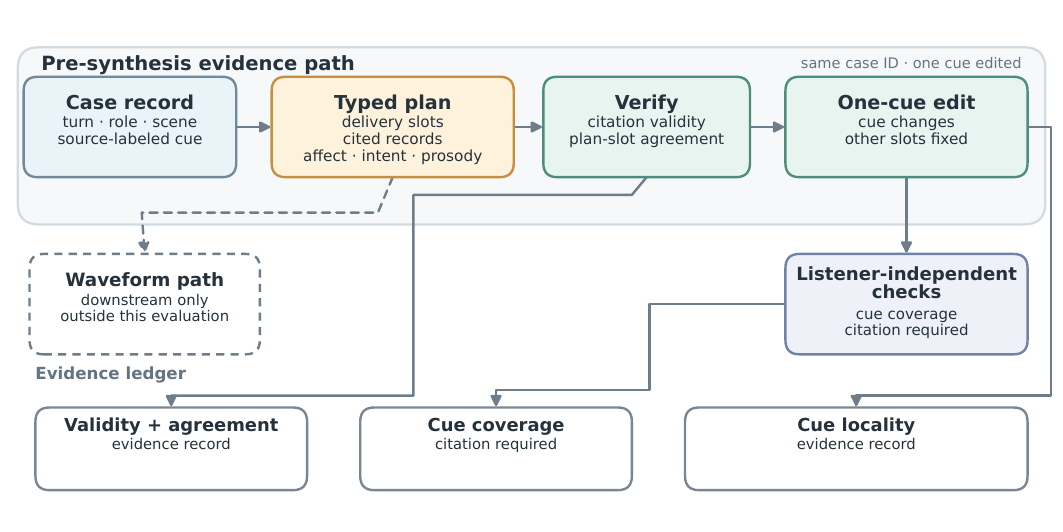}
\caption{VoxReason's pre-synthesis verification path. A case record is converted into a structured delivery plan, checked for citation validity and plan-slot agreement, and passed through a one-cue edit while the other delivery slots remain fixed. Automatic checks report cue coverage and citation requirements before synthesis. The lower ledger records validity and agreement, locality, and coverage and citation checks. The waveform path is downstream and outside the current structural evaluation. These checks do not measure listener judgments.}
\label{fig:overview}
\end{figure}

Figure~\ref{fig:overview} summarizes the verification path. A case record supplies the utterance context and source-label cue. A planner returns a schema-constrained delivery plan and cites the evidence it used. The verifier checks citation validity and slot agreement. A one-cue edit then changes the designated cue while keeping other slots fixed. The automatic checks report evidence coverage, citation-required grounding, and counterfactual locality. The waveform branch is outside the current evaluation.

\subsection{Metrics}

\begin{samepage}
For predicted evidence $\hat{C}$ and gold evidence $C^{\star}$, the verifier matches records by cue identifier and reports precision, recall, and F1. Let $m$ be the number of matched records. The formulas are
\begin{equation}
 P = \frac{m}{|\hat{C}|}, \qquad R = \frac{m}{|C^{\star}|}, \qquad F_1 = \frac{2PR}{P+R}.
\end{equation}
\end{samepage}
The decisive-cue recall metric restricts the recall calculation to the cue changed by the counterfactual edit. Hallucinated evidence rate is $1-P$ when evidence is predicted, and uncited evidence rate is $1-R$ when gold evidence exists.

Plan-slot accuracy is the mean agreement over the seven scalar fields and the exact set match for emphasis:
\begin{equation}
 A_{\mathrm{plan}} = \frac{\sum_{j=1}^{7} \mathbf{1}[\hat{p}_j=p^{\star}_j] + \mathbf{1}[\hat{p}_{\mathrm{emph}}=p^{\star}_{\mathrm{emph}}]}{8}.
\end{equation}
The released gold plan is a task specification; exact plan agreement does not establish perceptual appropriateness.
The verifier's grounded score combines evidence F1, plan-slot accuracy, and the absence of hallucinated evidence:
\begin{equation}
 G = 0.45F_1 + 0.45A_{\mathrm{plan}} + 0.10(1-H),
\end{equation}
where $H$ is the hallucinated evidence rate. The citation-required score multiplies $G$ by evidence recall, so missing required evidence lowers the score:
\begin{equation}
G_{\mathrm{cite}} = G R.
\end{equation}
This product lowers the grounded score when a plan is plausible but its supporting source records are missing.
For a counterfactual prediction, $G_{\mathrm{cite}}^{\mathrm{cf}}$ uses the same formula with that prediction's plan, evidence set, and counterfactual gold cues.
When predicted or gold evidence is empty, undefined precision or recall is set to zero, and F1 is zero when $P+R=0$. Hallucinated-evidence and uncited-evidence rates are zero when their respective reference sets are empty.

For each original-counterfactual pair, a required field earns credit only if it changes from the original prediction and reaches the declared value; a field that should remain fixed earns preservation credit when it is unchanged. Counterfactual consistency is the per-pair product of required-change accuracy and preservation rate, averaged across valid pairs. Unexpected-change rate is one minus preservation rate.

\section{Evaluation Protocol}

\paragraph{Study map and units.}
All three partitions in Table~\ref{tab:split-construction} rearrange the \emph{same} 100 source-label cases from two source actors; their test sets contain 16, 24, and 32 cases, respectively. The public, source-key-disjoint, and source-emotion-disjoint test sets overlap pairwise in 5, 4, and 8 case identifiers, and one case appears in all three. They are alternative diagnostics, not three independent cohorts or speaker-held-out replications. A case is the scoring unit for original-plan metrics. Counterfactual locality uses an original--edited pair as its unit: the source-key test set supplies 24 such pairs in each of five deterministic conditions, or 120 condition records on 24 shared pairs. The three held-out emotion-intensity source groups, not 120 independent examples, are the resampling units for the descriptive intervals. The source-emotion split holds out emotion labels; neither holdout excludes the two source actors from training. Appendix~\ref{app:learned-model-scope} explains why the separate learned-planner studies do not support a model comparison.

\subsection{Controlled information comparison}

The primary comparison uses all 100 cases in the released source-label measurement; the 67/17/16 split remains the package's train/development/test organization. The text-only control receives the neutral textual context without the source-label evidence needed to identify the designated cue. The gold-copy reference receives the released gold evidence and plan; it is a deterministic verifier reference, not a learned model. Both systems are scored by the same verifier. The comparison measures the information available in the source-label task and checks that the verifier responds to it as intended.

The evidence-sensitivity matrix uses the 60/16/24 source-key-disjoint split and evaluates each of five conditions once. The conditions are a gold-copy reference, a source-emotion prior, a text-and-role lookup, a citation-echo control, and a declared-edit positive control. The prior predicts the most frequent gold plan among training cases with the same source-emotion label; tied plans are resolved by choosing the lexicographically smallest serialized plan. The text-and-role lookup cites only textual cues, and citation echo substitutes an incorrect cue identifier. The positive control applies the declared edit directly to the lookup plan, so it tests whether the verifier detects the expected transition, not whether a planner can infer the edit. The expected deltas are identical across this split, so a shuffled-delta control would not create a distinct comparison and is omitted.

For locality metrics, descriptive 95\% intervals resample the three held-out source groups 10,000 times with fixed seed 20260911. They describe this fixed split and do not estimate uncertainty across new groups.

\subsection{Anti-shortcut checks}

The first check rebuilds the lookup after removing each test key from training. It tests whether the released split's perfect score depends on a direct key-to-plan mapping. A second check uses the same 100 cases in a 60/16/24 partition with no source key shared between training and test. The prior predicts from source emotion alone, without case-specific information or a citation target. Its scores are calibration diagnostics, not benchmark performance; this check asks whether an emotion-based plan responds to the designated cue edit.

Table~\ref{tab:split-construction} specifies the held-out boundary used for each split diagnostic.

\begin{table}[t]
\centering
\caption{Split construction for the reported source-label measurements. A source key is an emotion-intensity pair. The public split is the released benchmark partition; the two held-out splits are fixed diagnostic partitions.}
\label{tab:split-construction}
\small
\begin{tabularx}{\columnwidth}{@{}XrrrX@{}}
\toprule
\textbf{Split} & \textbf{Train} & \textbf{Dev} & \textbf{Test} & \textbf{Held-out boundary}\\
\midrule
Public source-label & 67 & 17 & 16 & Released case partition\\
Source-key-disjoint & 60 & 16 & 24 & No test emotion-intensity key appears in train\\
Source-emotion-disjoint & 56 & 12 & 32 & No test emotion appears in train\\
\bottomrule
\end{tabularx}
\end{table}

\section{Results}

\subsection{The text-only control misses the decisive source cue}

Table~\ref{tab:main-results} reports the paired source-label comparison. The text-only control obtains evidence F1 of $0.857\uparrow$, but its decisive-cue recall is $0.000\uparrow$. It achieves plan-slot accuracy $0.185\uparrow$ and citation-required grounded score $0.427\uparrow$. The gold-copy reference reaches $1.000\uparrow$ on every positive metric and $0.000\downarrow$ on hallucinated and uncited evidence by construction. These references show that the verifier distinguishes a control without the designated source record from the supplied gold evidence and plan. They do not show that a learned planner reaches the gold-copy reference.

\begin{table}[t]
\centering
\caption{Paired public source-label measurement over 100 cases. The gold-copy reference uses released gold evidence and plans to test verifier behavior; it does not measure model performance. Gold plans are task specifications, not perceptual targets. Higher is better for arrows up; lower is better for arrows down.}
\label{tab:main-results}
\small
\begin{tabularx}{\linewidth}{@{}Yrrrrrr@{}}
\toprule
{\bfseries\boldmath System} & {\bfseries\boldmath Ev. F1 $\uparrow$} & {\bfseries\boldmath Decisive $\uparrow$} & {\bfseries\boldmath Plan $\uparrow$} & {\bfseries\boldmath $G_{\mathrm{cite}}\uparrow$} & {\bfseries\boldmath Halluc. $\downarrow$} & {\bfseries\boldmath Uncited $\downarrow$}\\
\midrule
Text-only control & 0.857 & 0.000 & 0.185 & 0.427 & 0.000 & 0.250\\
\rowcolor{tablereference}
Source-label oracle & 1.000 & 1.000 & 1.000 & 1.000 & 0.000 & 0.000\\
\bottomrule
\end{tabularx}
\end{table}

The text-only control's $0.000\downarrow$ hallucinated-evidence rate should be read together with its $0.250\downarrow$ uncited-evidence rate. When it predicts evidence, it predicts only records allowed by the case, but it omits one quarter of the gold evidence on average. The citation-required score multiplies grounded quality by evidence recall and therefore records this omission. We use $G_{\mathrm{cite}}$ as the primary aggregate for source-grounded planning.

\subsection{Evidence sensitivity under a source-key holdout}

The controlled matrix in Table~\ref{tab:evidence-sensitivity} evaluates five conditions once on the same 24 source-key-disjoint pairs. Each condition is deterministic, so repeated seeds would repeat the same predictions. The gold-copy reference passes each verifier component; the text evidence plus declared edit control passes the locality checks because it receives the expected edit directly. This is a scorer sensitivity check, not evidence that a learned planner can infer the edit. Appendix~\ref{app:matrix-locality} reports the remaining locality metrics and descriptive intervals; Table~\ref{tab:matrix-evidence} reports the evidence components omitted from the main table.

The text evidence-only and citation-echo controls preserve all unaffected slots but have decisive-cue recall $0.000\uparrow$ and required-change accuracy $0.000\uparrow$. The prior-only control reaches original plan-slot accuracy $0.958\uparrow$, yet has counterfactual consistency $0.000\uparrow$ and required-change accuracy $0.000\uparrow$; it preserves $0.667\uparrow$ of unaffected slots and changes $0.333\downarrow$ unexpectedly. Under this verifier, a successful response must cite the decisive cue, change the required fields, and preserve the others. The shuffled-delta fixture is omitted because all expected edits in this split are identical, so shuffling does not create a distinct control.

\begin{table}[h]
\centering
\caption{Evidence-sensitivity matrix on the source-key-disjoint holdout. Each condition is evaluated once over the same 24 held-out pairs. The table reports the core evidence metrics; Appendix~\ref{app:matrix-locality} reports complementary locality outcomes. The gold-copy reference copies the released evidence and plans; the other conditions are deterministic controls or a prior-only calibration diagnostic, not learned-model results.}
\label{tab:evidence-sensitivity}
\scriptsize
\setlength{\tabcolsep}{1.8pt}
\resizebox{0.99\linewidth}{!}{%
\begin{tabular}{@{}p{0.27\linewidth}*{7}{>{\centering\arraybackslash}c}@{}}
\toprule
{\bfseries\boldmath Condition} & {\bfseries\boldmath N} & \shortstack{\bfseries Orig. plan\\\bfseries $\uparrow$} & \shortstack{\bfseries CF plan\\\bfseries $\uparrow$} & \shortstack{\bfseries Decisive\\\bfseries $\uparrow$} & \shortstack{\bfseries CF cons.\\\bfseries $\uparrow$} & \shortstack{\bfseries Req. change\\\bfseries $\uparrow$} & \shortstack{\bfseries\boldmath $G_{\mathrm{cite}}^{\mathrm{cf}}$\\\bfseries $\uparrow$}\\
\midrule
\rowcolor{tablereference}
Source-label oracle & 24 & 1.000 & 1.000 & 1.000 & 1.000 & 1.000 & 1.000\\
\rowcolor{tablediagnostic}
Text evidence + declared delta & 24 & 0.583 & 0.792 & 0.000 & 1.000 & 1.000 & 0.000\\
Text evidence only & 24 & 0.583 & 0.417 & 0.000 & 0.000 & 0.000 & 0.505\\
Citation echo control & 24 & 0.583 & 0.417 & 0.000 & 0.000 & 0.000 & 0.117\\
Prior only & 24 & 0.958 & 0.417 & 0.000 & 0.000 & 0.000 & 0.000\\
\bottomrule
\end{tabular}%
}
\vspace{-2pt}
\begin{flushleft}
\footnotesize ``Plan$_\mathrm{orig}$'' and ``Plan$_\mathrm{cf}$'' denote original and counterfactual plan-slot accuracy; ``CF cons.'' denotes counterfactual consistency; ``Req. change'' denotes required-change accuracy; $G_{\mathrm{cite}}^{\mathrm{cf}}$ is the citation-required grounded score computed on the counterfactual prediction.\end{flushleft}
\end{table}

\begin{figure}[t]
\centering
\includegraphics[width=\linewidth]{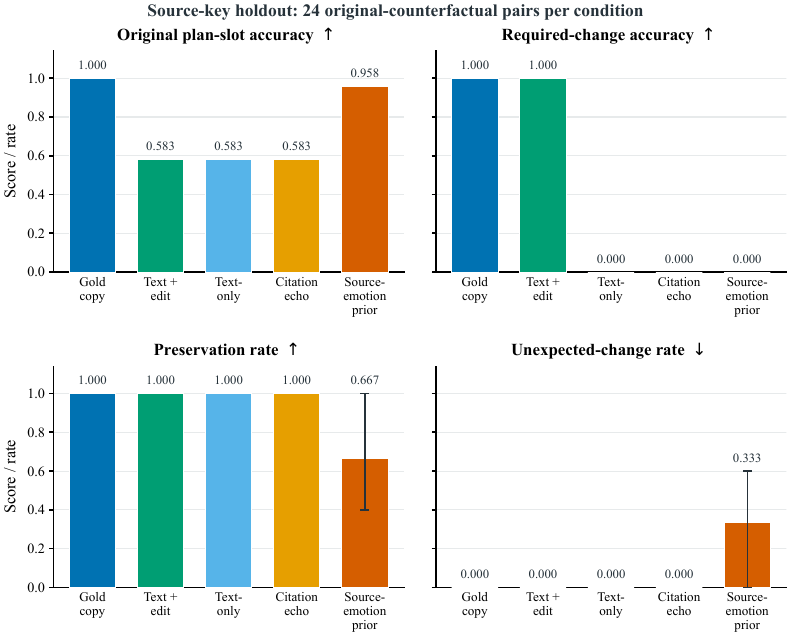}
\caption{Evidence-sensitivity results on the source-key-disjoint holdout. The five conditions are deterministic controls and a gold-copy verifier reference. Bars show results for 24 shared original-counterfactual pairs per condition. Whiskers appear only for prior-only preservation and unexpected-change rates; they are descriptive 95\% intervals from 10,000 resamples of the three observed source groups on this fixed split. The matrix tests verifier sensitivity and is not a learned-model comparison.}
\label{fig:evidence-sensitivity}
\end{figure}

Figure~\ref{fig:evidence-sensitivity} shows the separation between original plan agreement and response to the edit. The prior-only control has original plan-slot accuracy $0.958\uparrow$ but required-change accuracy $0.000\uparrow$; the gold-copy reference and declared-edit control pass the locality checks for different reasons.

\subsection{Prior-only diagnostics depend on label support}

Table~\ref{tab:shortcut} summarizes three fixed partitions of the same 100 cases, not independent cohorts. The prior predicts a plan from source emotion alone, so its training support for both original and edited labels matters.

The source-key split shares no emotion--intensity pair between training and test, but all 24 original test emotions appear in training and none of the 24 edited neutral targets does. The prior attains plan-slot accuracy $0.958\uparrow$, required-change accuracy $0.000\uparrow$, preservation $0.667\uparrow$, and unexpected-change rate $0.333\downarrow$. Because the edited targets are out of support, this result does not establish that the prior fails on supported emotion edits.

The source-emotion split holds out both test emotions (angry and calm), while all 32 edited neutral targets appear in training. The prior reaches plan-slot accuracy $0.219\uparrow$ and required-change accuracy $1.000\uparrow$, while preserving $0.400\uparrow$ of unaffected slots. In the public split, all 16 original emotions and edited targets appear in training; required-change accuracy is $1.000\uparrow$, but preservation is $0.200\uparrow$. These diagnostics show that original-plan agreement and local response measure different behavior, and that intervention scores depend on support for both original and target labels. Table~\ref{tab:evidence-sensitivity} provides the complementary verifier-sensitivity check on the source-key pairs.

\begin{table}[t]
\centering
\caption{Prior-only diagnostics on three fixed partitions of the same 100-case pool. The prior sees source emotion only and provides neither case-specific records nor citations. Original/target support for this lookup field is, respectively: public, 16/16 and 16/16; source-key-disjoint, 24/24 and 0/24; source-emotion-disjoint, 0/32 and 32/32. Counterfactual consistency combines required-change accuracy and preservation, so the component rates are shown separately.}
\label{tab:shortcut}
\small
\begin{tabularx}{\linewidth}{@{}YrrrrrX@{}}
\toprule
{\bfseries\boldmath Split} & {\bfseries\boldmath $N$} & {\bfseries\boldmath Plan-slot $\uparrow$} & \shortstack{\bfseries Required\\\bfseries change $\uparrow$} & {\bfseries\boldmath Preserve $\uparrow$} & \shortstack{\bfseries Unexpected\\\bfseries change $\downarrow$} & {\bfseries\boldmath Citations}\\
\midrule
Public source-label & 16 & 0.953 & 1.000 & 0.200 & 0.800 & None\\
Source-key-disjoint & 24 & 0.958 & 0.000 & 0.667 & 0.333 & None\\
Source-emotion-disjoint & 32 & 0.219 & 1.000 & 0.400 & 0.600 & None\\
\bottomrule
\end{tabularx}
\end{table}

Removing each exact source key from the lookup before scoring gives exact-plan accuracy $0.000\uparrow$ and plan-slot accuracy $0.266\uparrow$. This confirms that the perfect lookup score on the released split is a property of the released mapping, not evidence that the task is intrinsically solved without source grounding.

\subsection{What the current package does and does not establish}

The results support three limited conclusions. First, the public case construction passes deterministic checks: every released gold plan is enumerated against the prompt taxonomy, and each of the 15 emotion-intensity keys resolves to its declared plan mapping. Second, the verifier distinguishes legal evidence from missing evidence and rewards the expected plan changes. Third, plan-slot accuracy alone does not describe a prior's response to an edit: its intervention result depends on whether the prior has seen both the original emotion and the edited target during training. These conclusions concern a planning-stage measurement protocol. They do not assess speech naturalness, listener preference, speaker similarity, intelligibility, or generalization to unconstrained conversational audio.

\section{Discussion}

VoxReason treats grounding as a relation among a cue, a plan field, and a cited source record. That relation is easy to lose when evaluation collapses all quality into a single waveform score. The controlled edit makes the relation observable: because the utterance remains fixed, a plan change can be attributed to the edited cue. The cited source records let readers inspect this relation after scoring.

The benchmark's narrowness is a feature for this purpose. A small source-label suite can be checked line by line, regenerated without private audio, and used to debug output schemas before adding acoustic inputs. As a controlled diagnostic, it separates planning errors from rendering errors within the released source-label setting. The same narrowness also limits what the result can establish. Two target utterances, one scene label, and two roles are not enough to characterize context-aware speech planning in the wild. The benchmark is a controlled diagnostic, not a replacement for audio evaluation.

The prior-only diagnostics suggest a practical evaluation rule. Plan-slot accuracy does not show whether a predictor changes the fields required by an intervention or preserves the others. For label-based priors, reports should state training support for both the original input label and the edited target label; otherwise a failure to change may combine shortcut behavior with an unsupported target.

\section{Limitations and Responsible Use}

\paragraph{Data scope.}
The released benchmark is derived from RAVDESS labels and contains no public context-audio records. It therefore measures source-label planning, not audio-conditioned planning. The 100 clips come from two source actors, so they do not support speaker-level generalization. RAVDESS contains acted emotional speech; the derived cases are not a representative sample of natural conversation.

\paragraph{Construct scope.}
The suite has only two target utterances, one scene, two roles, and 15 source keys. The deterministic key-to-plan mapping is useful for validating the verifier but makes the released split vulnerable to memorization. The source-key split is a fixed calibration diagnostic, and its edited neutral targets are not supported by the prior's training emotions; it is not a clean in-support test of intervention response or a broad benchmark.

\paragraph{Metric scope.}
The verifier uses exact structured-field agreement and exact evidence matching. It does not judge whether a different but semantically reasonable plan would sound good, and it does not infer prosody from a waveform. The weighted grounded score is an engineering metric defined by the public implementation, not a human-validated perceptual scale.

\paragraph{Human and audio evaluation.}
We report no listener judgments, so perceived naturalness, appropriateness, and preference remain unmeasured. Assessing these outcomes requires a separately designed human-listener study with appropriate consent and data handling; listener judgments should remain distinct from verifier scores.

\section{Conclusion}

VoxReason measures a specific question: when a contextual cue changes, does a speech planner cite the source record that supports its decision and update only the linked delivery fields? The public source-label suite and verifier make that question measurable before synthesis. The controlled comparison shows the information difference between a text-only control and a gold-copy reference, while the shortcut checks show why plan-slot accuracy alone is insufficient. The results do not address listener experience or waveform quality. Future work should add audio-context inputs and human-listener evaluation of perceived naturalness, appropriateness, and preference.

\section*{Reproducibility Statement}
Sections 2--5 define the plan schema, metrics, controls, and deterministic measurements. The appendix gives the split diagnostics and extended results; the repository contains the derived source-label files, split builders, scorer, and result scripts. The release excludes source audio and does not require model inference or speech generation.

\clearpage
\section*{AI Use Statement}
Yes, to aid or polish writing. Details are described in the paper.\\
Yes, for retrieval and discovery (e.g., finding related work). Details are described in the paper.

\clearpage
\appendix
\section*{Appendix}
\FloatBarrier
\section{Appendix Roadmap}
This appendix supports the planning-stage measurements in Sections 3--5. Section~\ref{app:study-map} defines the populations, conditions, and units. Sections~\ref{app:matrix-locality} and \ref{app:split-details} report the counterfactual results and split diagnostics. Section~\ref{app:learned-model-scope} explains why separate learned-planner studies do not support a model comparison, and Section~\ref{app:acoustic-consistency} reports descriptive checks on derived acoustic features. These analyses do not measure listener judgments or generated-speech quality.

\section{Study Identity and Interpretation}\label{app:study-map}
The released 100-case pool is reused across all three partitions. Train, development, and test identifiers are disjoint within each partition, but test cases overlap across partitions. The public comparison scores both deterministic controls on all 100 cases; the prior diagnostic in Table~\ref{tab:shortcut} uses the 16 public test cases. The source-key test has 24 original--edited pairs, each evaluated under five matrix conditions, for 120 condition records. The source-emotion test has 32 cases. All cases come from two actors, so neither held-out partition is a speaker-held-out or independent-dataset estimate.

Training support differs across partitions. In the source-key test, all 24 original emotions appear in training but none of the 24 edited neutral targets does. In the source-emotion test, neither of the 32 original emotions appears in training, while all edited neutral targets do. The public test includes all 16 original emotions and edited targets. These differences affect how to interpret the prior's plan and intervention scores.

The automatic scorer matches cited cue identifiers, compares eight structured plan slots, and checks required changes and preservation under the designated edit. The source-label oracle copies supplied gold evidence and plans; the declared-edit control also receives the expected transition. Their perfect scores show that the scorer recognizes its task specification, not that an inferred plan or waveform is perceptually appropriate.

The prior and text-only controls remove different information. The source-emotion prior uses a training-derived label prior but has neither case-specific citations nor the edited cue. The text-only control receives textual context but lacks the decisive source label. Sections~\ref{app:matrix-locality} and \ref{app:split-details} report their outcomes and denominators.

\section{Expanded Evidence-Sensitivity Matrix}\label{app:matrix-locality}

Table~\ref{tab:matrix-locality} gives the locality outcomes for the same 24 pairs per condition. Counterfactual consistency is $1.000\uparrow$ for the gold-copy and declared-edit controls, and $0.000\uparrow$ for the other three conditions. The source-group bootstrap intervals resample the three groups — \texttt{happy:strong}, \texttt{angry:strong}, and \texttt{fearful:normal} — and describe this fixed split; they are not population-level uncertainty estimates. Only the prior's preservation rate ($0.667\uparrow$ [0.400, 1.000]) and unexpected-change rate ($0.333\downarrow$ [0.000, 0.600]) have nondegenerate intervals. The declared-edit control receives the expected edit directly, so its locality scores test verifier behavior, not a planner's ability to identify a source cue.

\begin{table}[t]
\centering
\caption{Locality outcomes for the source-key-disjoint evidence-sensitivity matrix. Each condition is evaluated once over 24 pairs. ``Required'' is required-change accuracy, ``preserve'' is the rate of keeping unaffected slots unchanged, and ``unexpected'' is the rate of changing an unaffected slot.}
\label{tab:matrix-locality}
\small
\begin{tabularx}{\columnwidth}{@{}Xrrr@{}}
\toprule
\textbf{Condition} & \textbf{Required $\uparrow$} & \textbf{Preserve $\uparrow$} & \textbf{Unexpected $\downarrow$}\\
\midrule
Source-label oracle & 1.000 & 1.000 & 0.000\\
Text evidence + declared delta & 1.000 & 1.000 & 0.000\\
Text evidence only & 0.000 & 1.000 & 0.000\\
Citation echo control & 0.000 & 1.000 & 0.000\\
Prior only & 0.000 & 0.667 & 0.333\\
\bottomrule
\end{tabularx}
\vspace{-2pt}
\end{table}

Table~\ref{tab:matrix-evidence} separates evidence matching from plan behavior. Text evidence only achieves evidence F1 $0.857\uparrow$ but has decisive-cue recall $0.000\uparrow$ and required-change accuracy $0.000\uparrow$; citation echo provides precision without the decisive cue. Conversely, the prior has original plan-slot accuracy $0.958\uparrow$ without evidence. These controls show that neither a plausible citation nor static plan agreement substitutes for a source-grounded response to the edit.

\begin{table}[t]
\centering
\caption{Evidence diagnostics for the original predictions in the source-key-disjoint matrix. Each condition is evaluated over 24 pairs. ``Uncited'' is the rate of required evidence that is not cited; the hallucinated-evidence rate is $0.000\downarrow$ for every condition.}
\label{tab:matrix-evidence}
\small
\begin{tabularx}{\columnwidth}{@{}Xrrrr@{}}
\toprule
\textbf{Condition} & \textbf{Evidence F1 $\uparrow$} & \textbf{Decisive recall $\uparrow$} & \textbf{Uncited $\downarrow$} & \textbf{$G_{\mathrm{cite}}$ $\uparrow$}\\
\midrule
Source-label oracle & 1.000 & 1.000 & 0.000 & 1.000\\
Text evidence + declared delta & 0.000 & 0.000 & 1.000 & 0.000\\
Text evidence only & 0.857 & 0.000 & 0.250 & 0.561\\
Citation echo control & 0.400 & 0.000 & 0.750 & 0.136\\
Prior only & 0.000 & 0.000 & 1.000 & 0.000\\
\bottomrule
\end{tabularx}
\end{table}

\FloatBarrier
\section{Split and Anti-Shortcut Details}\label{app:split-details}
The three partitions are defined in Table~\ref{tab:split-construction} in Section~4. They are fixed partitions, not repeated samples from a population of speakers or conversations.

Section~5 reports the lookup result after removing each test key from the training mapping. This tests dependence on exact keys in the released suite, not the behavior of a learned planner trained on broader evidence. The prior result for the partition that holds out source emotion labels is likewise a narrow shortcut diagnostic (Table~\ref{tab:shortcut}).

\FloatBarrier
\section{Scope of Auxiliary Learned-Model Studies}\label{app:learned-model-scope}
The results in this paper are deterministic verifier diagnostics. Separate studies of learned planners do not support a model ranking or a matched performance effect because their target definitions and evaluation protocols differed. Changing the expected counterfactual plan changes the locality score even when predictions are unchanged. A valid comparison would require the same cases, fixed target definitions, a shared scorer and input protocol, and predictions for every case.

Separate RAVDESS and CREMA-D source-label reports use oracle copies of supplied labels and plans. They do not evaluate learned planning, audio input, or transfer to new speakers and therefore do not extend the results reported here.

\section{Acoustic Feature Consistency}\label{app:acoustic-consistency}

The acoustic feature summary covers 100 unique source-label cases. The resulting means are 3.770 seconds for duration, 0.570 for silence fraction, 0.349 for voiced fraction, 207 Hz for rough pitch, and 0.063 for zero-crossing rate. Table~\ref{tab:acoustic-consistency} compares these source labels with root-mean-square (RMS), peak amplitude, rough pitch, plan pitch, plan energy, and zero-crossing rate. The 100 cases come from two source actors. Both actors show a positive mean difference on each comparison, so we report the pooled summaries descriptively and do not estimate uncertainty across speakers. These measurements check consistency between source labels and extracted features; they are not audio-input, synthesis-quality, or listener results.

\begin{table}[H]
\centering
\caption{Descriptive source-label and acoustic-feature comparisons over 100 clips from two actors. Area under the receiver operating characteristic curve (AUROC) compares positive and negative clips pooled across both actors; $\Delta$ is the positive-minus-negative mean in feature units. The mean difference is positive within each actor. These values do not estimate performance for other speakers and do not measure audio input, generated-speech quality, or listener judgments.}
\label{tab:acoustic-consistency}
\small
\begin{tabularx}{\columnwidth}{@{}Xrrr@{}}
\toprule
\textbf{Contrast (feature)} & \textbf{$N_+/N_-$} & \textbf{AUROC $\uparrow$} & \textbf{$\Delta$ $\uparrow$}\\
\midrule
Strong $-$ normal intensity (RMS) & 44/56 & 0.728 & 0.0179\\
Strong $-$ normal intensity (peak) & 44/56 & 0.716 & 0.180\\
Strong $-$ normal intensity (rough pitch, Hz) & 44/56 & 0.674 & 38.9\\
Raised $-$ lowered plan pitch (Hz) & 52/40 & 0.713 & 48.4\\
High $-$ low plan energy (RMS) & 60/24 & 0.788 & 0.0153\\
Fast $-$ slow plan rate (zero-crossing rate) & 36/32 & 0.769 & 0.0214\\
\bottomrule
\end{tabularx}
\end{table}

These measurements do not establish audio-conditioned planning, waveform quality, or listener preference. The source-label counterfactual matrix remains the central result; the acoustic checks do not change its conclusion.

\clearpage

\begin{thebibliography}{11}

\bibitem[Livingstone and Russo(2018)]{ravdess}
Steven R. Livingstone and Frank A. Russo.
\newblock The Ryerson Audio-Visual Database of Emotional Speech and Song (RAVDESS): A dynamic, multimodal set of facial and vocal expressions in North American English.
\newblock \emph{PLOS ONE}, 13(5):e0196391, 2018.
\newblock \url{https://doi.org/10.1371/journal.pone.0196391}.

\bibitem[Zhang et~al.(2023)]{speechgpt}
Dong Zhang, Shimin Li, Xin Zhang, Jun Zhan, Pengyu Wang, Yaqian Zhou, and Xipeng Qiu.
\newblock SpeechGPT: Empowering large language models with intrinsic cross-modal conversational abilities.
\newblock arXiv preprint arXiv:2305.11000, 2023.
\newblock \url{https://arxiv.org/abs/2305.11000}.

\bibitem[Huang et~al.(2023)]{audiogpt}
Rongjie Huang, Mingze Li, Dongchao Yang, Jiatong Shi, Xuankai Chang, Zhenhui Ye, Yuning Wu, Zhiqing Hong, Jiawei Huang, Jinglin Liu, Yi Ren, Zhou Zhao, and Shinji Watanabe.
\newblock AudioGPT: Understanding and generating speech, music, sound, and talking head.
\newblock arXiv preprint arXiv:2304.12995, 2023.
\newblock \url{https://arxiv.org/abs/2304.12995}.

\bibitem[Ma et~al.(2025)]{mmar}
Ziyang Ma, Yinghao Ma, Yanqiao Zhu, Chen Yang, Yi-Wen Chao, Ruiyang Xu, Wenxi Chen, Yuanzhe Chen, Zhuo Chen, Jian Cong, Kai Li, Keliang Li, Siyou Li, Xinfeng Li, Xiquan Li, Zheng Lian, Yuzhe Liang, Minghao Liu, Zhikang Niu, Tianrui Wang, Yuping Wang, Yuxuan Wang, Yihao Wu, Guanrou Yang, Jianwei Yu, Ruibin Yuan, Zhisheng Zheng, Ziya Zhou, Haina Zhu, Wei Xue, Emmanouil Benetos, Kai Yu, Eng-Siong Chng, and Xie Chen.
\newblock MMAR: A challenging benchmark for deep reasoning in speech, audio, music, and their mix.
\newblock arXiv preprint arXiv:2505.13032, 2025.
\newblock \url{https://arxiv.org/abs/2505.13032}.

\bibitem[Liu et~al.(2025)]{thinksound}
Huadai Liu, Kaicheng Luo, Jialei Wang, Wen Wang, Qian Chen, Zhou Zhao, and Wei Xue.
\newblock ThinkSound: Chain-of-thought reasoning in multimodal large language models for audio generation and editing.
\newblock In \emph{Advances in Neural Information Processing Systems}, 2025.
\newblock arXiv:2506.21448.

\bibitem[Xue et~al.(2026)]{cottts}
Wei Xue, Junlan Feng, Shilei Zhang, Yue Wang, Ruosong Yang, Bei Liu, Liumeng Xue, Sitong Cheng, Jiahao Pan, Weizhen Bian, Boyi Kang, and Bin Long.
\newblock ISCSLP 2026 CoT-TTS Challenge: Chain-of-thought reasoning for context-aware text-to-speech.
\newblock arXiv preprint arXiv:2606.21933, 2026.
\newblock \url{https://arxiv.org/abs/2606.21933}.

\bibitem[Ye et~al.(2024)]{codec}
Zhen Ye, Peiwen Sun, Jiahe Lei, Hongzhan Lin, Xu Tan, Zheqi Dai, Qiuqiang Kong, Jianyi Chen, Jiahao Pan, Qifeng Liu, Yike Guo, and Wei Xue.
\newblock Codec does matter: Exploring the semantic shortcoming of codec for audio language model.
\newblock arXiv preprint arXiv:2408.17175, 2024.
\newblock \url{https://arxiv.org/abs/2408.17175}.

\bibitem[Du et~al.(2024)]{unicats}
Chenpeng Du, Yiwei Guo, Feiyu Shen, Zhijun Liu, Zheng Liang, Xie Chen, Shuai Wang, Hui Zhang, and Kai Yu.
\newblock UniCATS: A unified context-aware text-to-speech framework with contextual VQ-diffusion and vocoding.
\newblock In \emph{Proceedings of the AAAI Conference on Artificial Intelligence}, volume 38, 2024.
\newblock \url{https://doi.org/10.1609/aaai.v38i16.29747}.

\end{thebibliography}
\end{document}